\documentclass[12pt,a4paper]{article}
\usepackage[utf8]{inputenc}
\usepackage{amsfonts}

\usepackage{epsf,array,delarray,amsmath,amssymb,amsthm}
\usepackage{graphicx}
\usepackage{cite}

\newcommand{\sM}{{\mathcal M}}
\newcommand{\sF}{{\mathcal F}}
\newcommand{\sS}{{\mathcal S}}

\newcommand{\Sig}{ \Sigma \mskip -8mu \vert \mskip 7mu}
\newcommand{\half}{  {\scriptstyle \frac{1}{2} }\,  }

\newcommand{\sX}{{\mathcal X}}
\newcommand{\R}{   {\mathbb R}  }
\newcommand{\TT}{{\mathbb T}}
\newcommand{ \Nbins}{ N_{ \rm bins } }
\newcommand{ \Nsim}{ N_{ \rm sim } }
\newcommand{\Nsub}{  N_{ \rm sub } }

\newcommand{\Nblock}{ N_{ \rm block}  }
\newcommand{ \Nprimal}{  N_{\rm primal} }
\newcommand{\Ndual}{  N_{\rm dual} }

\theoremstyle{plain}

\newtheorem{theorem}{Theorem}[section]

\newtheorem{definition}[theorem]{Definition}
\numberwithin{equation}{section}

\author{L.~C.~G. Rogers}
\title{Bermudan options by simulation}
\begin{document}
\maketitle
\begin{abstract}
The aim of this study is to devise numerical methods for dealing with very high-dimensional Bermudan-style derivatives. For such problems, we quickly see that we can at best hope for price bounds, and we can only use a simulation approach. We use the approach of Barraquand \& Martineau \cite{BM} which proposes that the reward process should be treated as if it were Markovian, and then uses this to generate a stopping rule and hence a lower bound on the price.  Using the dual approach introduced by Rogers \cite{R1} and Haugh \& Kogan  \cite{HK}, this approximate Markov process leads us to hedging strategies, and upper bounds on the price.  The methodology is generic, and is illustrated on eight examples of varying levels of difficulty. Run times are largely insensitive to dimension.
\end{abstract}

\section{Introduction.}\label{intro}
A general optimal stopping problem can be formulated as finding
\begin{equation}
\sup_{ 0 \leq \tau \leq T} E[ \, Z_\tau \, ]
\label{obj1}
\end{equation}
where the termination time $T$ is a fixed positive real, the process $Z$ is a given process adapted to some filtration $(\sF_t)_{t \geq 0}$, and $\tau$ is constrained to be an $(\sF_t)$-stopping time. We shall in this paper assume that the reward process $Z$ has the form
\begin{equation}
Z_t  = g(t, X_t),
\label{Zdef}
\end{equation}
where $X$ is some Markov process, and $g$ is some measurable function of time and $X_t$. For what mainly concerns this paper - the analysis of Bermudan options - this formulation is already sufficiently general. Much of what follows is valid more widely, but at times we shall refer to properties that relate to a typical financial context; the aim is to provide bounds, and we need context in order to assess the effectiveness of these bounds.

Of course, it is well known how to solve an optimal stopping problem of the form given by \eqref{obj1}, \eqref{Zdef}; we define the value function 
\begin{equation}
V^*(t,x) = \sup_{t \leq \tau \leq T} E[ \, g(\tau,X_\tau) \,
\vert \, X_t = x \, ]
\label{valdef}
\end{equation}
and find $V^*$ by dynamic programming\footnote{For a Bermudan option, which has only finitely many possible times of exercise, the optimization is over a discrete set of times, though we generally think that the time of the underlying process runs continuously.}.  When this can be done, this is everything we could ask for. However, explicitly-soluble examples are rare, and we soon have to reach beyond them. For example, the standard Bermudan put option, where 
\begin{equation}
g(t,x) =e^{-rt} (K - \exp(x) )^+
\label{BP1}
\end{equation}
and the process $X$ is a Brownian motion,  is a celebrated example where no closed-form solution is known. This has launched many studies in the last 25 years,  and very efficient numerical schemes have been devised.  But what if we have some vector $X_t = (X^1_t, \ldots, X^d_t)$ of correlated Brownian motions, and the reward function is 
\begin{equation}
g(t, x) = e^{-rt} (\, K - \exp(\, \min_i x^i\, )\, )^+ \quad ,
\label{BP2}
\end{equation}
a so-called {\em min put}?  If we look for the value function, we have to determine numerically\footnote{If we can only solve the one-asset problem numerically, we can certainly only solve the $d$-asset problem numerically.} some function of $d$ variables. The standard Bellman equation approach requires us to calculate recursively  $V^*_t(\cdot) \equiv V^*(t, \cdot)$ as
\begin{equation}
V^*_t(x) = \max\{ \; g(t,x),\; E[ V^*_{t+1}(X_{t+1}) \vert X_t = x]\; \}\;  ;
\label{bellman}
\end{equation}
but if $d$ is large (fifty, say), how is $V^*_{t+1}$ to be stored? How is the expectation on the right-hand side of \eqref{bellman} to be calculated or approximated?  The more one thinks about these issues, the clearer it becomes that {\em calculating an approximation to the value function can only work in dimensions that are not too high, and will most likely rely heavily on the structure of the problem under study.}  In other words, {\em any methodology that attempts to identify the value function will be of restricted applicability.}

\medskip
So we must be content with less; but less may be enough. If we had calculated the value function, what use would we make of it? We would use it to {\em determine the optimal control}: at each time $t$, we would see the state $X_t=x$ of the system, and if $V^*_t(x)> g(t,x)$ we would continue, otherwise we would stop.  We would use it to determine a {\em fair price } to pay or ask for the derivative at time 0. We would use it to {\em  delta-hedge} the derivative once it had been sold. The view taken in this paper is that the analysis of a Bermudan option requires just these:
\begin{itemize}
\item At each time, whatever the state, we are able to decide whether or not to stop;
\item At each time, we are able to propose a hedge for the next period;
\item We are able to provide reasonably close bounds for the price of the derivative at all times.
\end{itemize}
Exact knowledge of the value function would achieve all of these objectives, but can we attain them without knowing the value function or an approximation to the value function? The message of this paper is that this {\bf can} be done. We present an approach to Bermudan options with the properties:
\begin{itemize}
\item The {\em only} information required about the Markov process is the ability to simulate a step of the process;
\item The methodology is {\em generic} - the same code is used to do the calculations for all examples, changing only the specification of the Markov process $X$ and the stopping reward $g$;
\item Computational cost is {\em largely insensitive to dimension}, so a derivative written on hundreds of underlyings can be solved simply by changing the dimension parameter in the code, and takes only a little longer to run;
\item Upper and lower bounds on the price differ by typically 5-10 percent (sometimes more, sometimes less);
\item The stopping rules and hedging rules obtained are very simple to calculate and implement.
\item Computational times depend on the problem, but a few tens of seconds usually suffices.
\end{itemize}
The claim is that the method offered here is an effective general method for dealing with any Bermudan option. In truth, the components used are already known in one form or another, and what is added here is the judicious combination of them, and a redefinition of the questions to be answered. The waypoints are the following:
\begin{itemize}
\item Any numerical scheme has to be a finite calculation, so the Markov process has to be coerced to a finite set of values;
\item The finite coercion of the underlying Markov process $X$ has to be tailored to the stopping reward - using the approach of Barraquand \& Martineau \cite{BM};
\item The finite coercion generates a stopping rule, and a hedging rule, using the dual approach of \cite{R1}, \cite{HK}, \cite{AB} which can then be evaluated by simulation.
\end{itemize}
The particular structure of a problem may suggest variants that improve on the performance, but in the examples studied any improvement from simple variants is not large.

\section{The general methodology.}\label{S2}
The general situation concerns a Markov process $X$ taking values in a statespace $\sX$, a stopping reward function $g$, and a finite set $\TT \subseteq [0,T]$ of size $N_T$ of possible times to stop\footnote{We will always assume that 0 and $T$ are in $\TT$.}. The aim is to associate this problem with an optimal stopping problem for a (discrete-time) Markov process with a {\em finite} statespace, and the way this is done is simply a paraphrase of the method of Barraquand \& Martineau \cite{BM}.

\begin{definition}
A  (real-valued) {\em Markovian coercion} of $X$ is specified by a measurable function $g :\TT \times  \sX \mapsto \R$, an $N_T \times \Nbins$ matrix $\eta$ of {\em bin values} , an $N_T\times ( \Nbins-1)$ matrix $y$ of {\em bin edges}, and an $(N_T-1) \times \Nbins\times \Nbins$ array $P$ with the properties:
\begin{enumerate}
\item for each $t \in \TT$, 
\[
\eta_t^{(1)} < y_t^{(1)} < \eta_t^{(2)} < y_t^{(2)} < \ldots < y_t^{(\Nbins-1)} < \eta_t^{(\Nbins)}   ;
\]
\item  for each $t \in \TT\backslash\{T\}$, $P(t, \cdot, \cdot)$ is an $\Nbins \times \Nbins$ transition matrix.
\end{enumerate}
\end{definition}

We can approximate the real-valued process $g(t,X_t)$, $t \in \TT$, by using the matrix $y$ of bin edges to define a partition of $\R$ for each $t \in \TT$:
\begin{equation}
J_t^{(1)}=(-\infty, y_t^{(1)}]\; , \ldots,\;  J_t^{(N)} = ( y_t^{(N-1)}, \infty)
\end{equation}
and then the matrix $\eta$ of bin values to define the approximating process
\begin{equation}
Y_t  = \sum_{k=1}^N  \eta^{(k)}_t \; I_{ \{ g(t, X_t) \in J^{(k)}_t\}  }  \;  .
\label{Ydef}
\end{equation}
For the current purpose, we shall take $g$ to be the function appearing in  the definition \eqref{Zdef} of the stopping reward.

\medskip
The process $Y$ takes only finitely many values, but will {\em not} in general be Markov.   Nevertheless, the essential idea of the Barraquand-Martineau approach is that we {\bf pretend that it is}, with transition probabilities given by the array $P$.  We then solve the optimal stopping problem for this Markov coercion, and use the solution found to propose exercise and hedging strategies for $X$.

\subsection{Implementation.}\label{S2.1}
For the implementation, we shall assume that the state process $X$ takes values in some (subset of) euclidean space $\R^d$. The dimension $d$ of this space is unrestricted, and can be quite large. 
The numerical implementation consists of four stages:
\begin{enumerate}
\item use simulation to initialize, calculating the transition matrix array $P$, and the bin edges and values;
\item calculate the value and optimal stopping rule for the (presumed Markovian) process $Y$;
\item find a lower bound by evaluating the performance of the stopping rule from step 2 when applied to stopping the process $X$;
\item find an upper bound by evaluating the hedging rule derived from step 2 when applied to the actual process $X$.
\end{enumerate}
We now give some more detail on each of these stages in turn.

\medskip
\noindent{\bf Initialization.} First choose some number $\Nbins$ of bins, and some number $\Nblock$ of points in each `half bin', so that in total we will simulate $\Nsim = 2  \Nblock  \Nbins$ sample paths.  The simulation will fill up a $N_T \times d \times \Nsim$ array, though since the process is Markovian, we can economize on space by just carrying along the current values, in effect a $d \times \Nsim$ array, which gets overwritten with newly-calculated values.  

So suppose that we have the $d \times \Nsim$ array $x$ of the values\footnote{In accordance with Python notation, we denote the $(i,j)$ element of an array $z$ by $z[i,j]$, and the $j$th column of $z$ by $z[:,j]$.} at time $t_i \in \TT$. We now use the simulation of the Markov process to step these values forward to  the next time $t_{i+1}$, creating a $d \times \Nsim$ array $x'$. Next apply the function $g$ to give
\begin{equation}
y_j  \equiv g(t_i, x[:,j]), \qquad y'_j \equiv g(t_{i+1}, x'[:,j])
\qquad  (j = 1, \ldots, \Nsim).
\nonumber
\end{equation}
If we define $y_{(1)} < y_{(2)} < \ldots y_{(\Nsim)}$ to be the sample $(y_j)$ in increasing order, we define the bin edges at time $t_i$ to be the values $y_{(2k\Nblock)}$, $k=1, \ldots , \Nbins-1$, and the bin values at time $t_i$ to be the values $y_{((2k-1)\Nblock)}$, $k=1, \ldots , \Nbins$.  We similarly calculate\footnote{We implicitly assume that the $y$-values are distinct. This is clearly not going to happen for (say) a put option, but we make this happen by replacing the reward $\max\{ 0, K-S\} $ by the reward $\max \{ \varepsilon (K-S), (K-S) \}$ for some small $\varepsilon$. The error committed will be small compared to other errors. Likewise, we do not bother to locate the bin edges in between values of $y_{(j)}$ as perhaps we ought.} the bin values and edges at time $t_{i+1}$. Now for each simulated path $j$  we can see which bin that path was in at time $t_i$, and which bin that path moved into at time $t_{i+1}$; counting the number of paths which moved from bin $\ell$ at time $t_i$ into bin $m$ at time $t_{i+1}$ gives us an estimate of the transition probability $P[i,\ell, m]$.

It is worth remarking that this step differs slightly from what Barraquand \& Martineau do; we let the data tell us where the bin edges should be, and Barraquand \& Martineau set the bin edges before any simulation takes place,  as an {\it a priori} modelling choice. A similar use of the simulated data to determine an approximation procedure appears in Bouchard \& Warin \cite{bouchard}.

\medskip\noindent
{\bf Calculating the value function and optimal stopping rule.} Now that we have made a Markov chain proxy which jumps at discrete times from one bin to another according to the transition probabilities stored in $P$, the calculation of the value $V$ and stopping rule  $\sS$ is done by dynamic programming. The arrays $V$ and $\sS$ are both $N_T \times \Nbins$\, ; the values in $\sS$ are either 0 or 1, where $\sS[i,k]=1$ signifies that we should stop if at time $t_i$ the $Y$-value is in bin $k$. The time taken to compute $V$ and $\sS$ is negligible. As a notational convenience,  for $y \in \R$ we shall write
\begin{equation}
V(t,y) = V[i,k], \qquad \sS(t,y) = \sS[i,k]
\end{equation}
when $t = t_i \in \TT$  and $y \in J^{(k)}_t$, the $k$th bin at time $t_i$.

\medskip\noindent
{\bf Lower bound.} The computed stopping rule $\sS$ is now used to provide a lower bound for the value of the option. We simply simulate a large number of paths of the process $X$, and for each path we stop the first time $t_i$ for which $\sS(t_i,g(t_i, X_{t_i}))  = 1$. For each path, this gives a stopping value, and the lower bound is just the average over all paths.

\medskip\noindent
{\bf Upper bound.} To derive an upper bound, we need to recall some results about the dual approach from \cite{R1}, \cite{HK}, where it is shown that the value \eqref{obj1} of the optimal stopping problem has the alternative characterization
\begin{equation}
\sup_{ 0 \leq \tau \leq T} E[ \, Z_\tau \, ]
= \min_{M \in \sM_0} E [ \; \sup_{0 \leq t \leq T}\, (Z_t-M_t) \; ],
\label{obj2}
\end{equation}
as a minimum over the space $\sM_0$  of martingales vanishing at 0. The interpretation of any $M \in \sM_0$ as a hedging martingale is explained in \cite{R1}; this is an important part of the approach adopted here, because {\em the approximate hedge is achieved by a `good' martingale from the dual approach, not by delta-hedging}. As a consequence, it is not essential that we can evaluate an option price to high accuracy; the main driver for demanding high accuracy in pricing is to perform delta hedging, yet the present approach completely steps around all such considerations. Indeed, the entire method presented here works for {\em any} Markov process $X$, including finite-state Markov chains, for which the concept of delta hedging is as meaningless as differentiating with respect to an integer argument.

 The minimum on the right-hand side is attained when $M$ is the martingale part of the Snell envelope process: see \cite{R1}. In this setting, the Snell envelope process is simply the value function $V^*(t, X_t)$ evaluated along the path, so the optimal martingale difference sequence would be
\begin{equation}
 M^*_{t_{i+1}}  - M^*_{t_i} = V^*(t_{i+1}, X_{t_{i+1}}) - E[
 V^*(t_{i+1}, X_{t_{i+1}})\; \vert \; \sF_{t_i}].
 \label{mgdiff1}
\end{equation}
Of course, we do not know $V^*$, but we have calculated and stored in the array $V$ some approximation to $V^*$, so a natural approximation to $M^*$ would be found by taking the martingale difference sequence
\begin{equation}
 M_{t_{i+1}}  - M_{t_i} = V(t_{i+1}, Y_{t_{i+1}}) - E[
 V(t_{i+1}, Y_{t_{i+1}})\; \vert \; \sF_{t_i}],
 \label{mgdiff2}
\end{equation}
where $Y$ is as defined at \eqref{Ydef}. This is essentially the approach of Andersen \& Broadie \cite{AB}. The only issue with this is how we are to evaluate the conditional expectation on the right-hand side of \eqref{mgdiff2}. The function $V(t_{i+1}, \cdot)$ is a simple function, taking only $\Nbins$ values, but if we have simulated a sample path and we see $X_{t_i}=x$, how are we to calculate (or approximate) $E[V(t_{i+1}, Y_{t_{i+1}})\; \vert \;X_{t_i}=x\;]$?  The approach adopted is to perform a subsimulation of some $\Nsub$ values of $X_{t_{i+1}}$ starting from $X_{t_i}=x$. It is generally considered a bad idea to perform subsimulations, because this will usually take a lot of time, and may not be very accurate, but in this application the approach is effective {\em because $Y_{t_{i+1}}$ is scalar.} The importance of this is that our subsimulations do not need to search out some high-dimensional space, they only need to search out the real line; in practice, if the payoff $g$ is continuous, and the time-step not too large, most values of $g(t_{i+1}, X_{t_{i+1}})$ will be fairly close to $g(t_i, X_{t_i})$, so a relatively small number of subsimulations will suffice. In the examples reported later, we used only a few tens of subsimulations, usually on 4000 simulated paths; full details are reported for each example studied.

This explains how we construct candidate hedging martingales from the array $V$. In a financial context, we would want to be able to express such martingales in terms of traded assets; this could be done by calculating delta-hedges for the martingale differences arising from the approximation, but since this would be application-specific, we prefer not to go into the detail of how this would be done. The approach offered merely indicates a martingale to be used for hedging, not how exactly this is to be synthesised from marketed assets.


\section{Examples.}\label{S3}
Here we present a range of examples to illustrate the methodology, some familiar from the literature, others not.  In the following examples, we will be considering a $d$-vector of log-Brownian assets whose prices $S^i_t$ at time $t$ evolve as
\begin{equation}
dS^i_t  = S^i_t \, \biggl(  \; \sum_{j=1}^d \sigma_{ij}\, dW^j_t + \mu_i\, dt\; \biggr),
\label{dS}
\end{equation}
where $W$ is a $d$-dimensional Brownian motion, the $\sigma_{ij}$ and $\mu_i$ are previsible processes which are assumed {\em constant}, except in Example \ref{SISV}.  We write 
\begin{equation}
 \Sig \equiv \sigma \sigma^T,
 \label{Sigdef}
\end{equation}
a positive-definite symmetric matrix.  Since the focus is on derivative pricing, we shall assume that we are working in the risk-neutral measure, which amounts to the condition
\begin{equation}
\mu_i = r - \half \Sig_{ii},
\label{mudef}
\end{equation}
where $r$ is the riskless rate of interest, assumed {\em constant}, except in Example \eqref{SISV}. 

It proves convenient in most of the examples to simulate the discounted log prices
\begin{equation}
x^i_t \equiv -\int_0^t r_s \; ds + \log S^i_t = \sum_j \sigma_{ij} W^j_t - \half\Sigma_{ii} \, t  + \log S^i_0.
\label{xdef}
\end{equation}

\medskip\noindent
{\bf Variants.}  If $N$ is a positive martingale, $N_0=1$, then 
\begin{eqnarray*}
\sup_{0 \leq \tau \leq T} E\biggl[ \; g(\tau, X_\tau) \; \biggr]
&=& \sup_{0 \leq \tau \leq T} E\biggl[ N_\tau\;  \frac{\; g(\tau, X_\tau)}
{N_\tau} \;\biggr ]
\\
&=& \sup_{0 \leq \tau \leq T} E\biggl[ N_T \;\frac{\; g(\tau, X_\tau)}
{N_\tau} \; \biggr]
\\
&=&  \sup_{0 \leq \tau \leq T} \tilde E\biggl[ \frac{\; g(\tau, X_\tau)}
{N_\tau} \;\biggr ],
\end{eqnarray*}
where 
\begin{equation}
\frac{d\tilde P}{dP} = N_T.
\label{Ptildedef}
\end{equation}
If the martingale $N$  can be expressed as $N_t = \psi(t,X_t)$, then  we can work in the new measure $\tilde P$ with the new reward $\tilde g(t,x) = g(t,x)/\psi(t,x)$, and it may be convenient sometimes to do this.  This change of numeraire approach is reminiscent of Jamshidian's \cite{J1} version of dual American option valuation.  We could similarly transform the stopping reward $g(t,X_t)$ to $g(t,X_t) - \varphi(t,X_t) + \varphi(0,X_0)$, where $\varphi$ is some function for which $\varphi(t,X_t)$ is a martingale, for example, the value of the European option. 

\bigbreak
In the examples which follow, we calculate  the simulation-based upper and lower bounds for various derivative prices. In the tables presented, we report also the simulation values of the European equivalent derivative. This should of course always be cheaper than the Bermudan option. In some places this natural inequality appears to be violated by a small amount. Partly this is because the standard error inherent in the simulation method (reported in parentheses after each entry in the tables), but also because the European prices are computed from the averages of the {\em actual} values of the option at expiry, whereas the lower bounds are based on the discretized values of the the option at expiry, and these are not the same. Discrepancies are in any case all very small even when present.

The total compute time is reported also; usually, the main part of the calculation was the calculation of the dual upper bound. Sometimes the initial calculation of the transition matrices was quite time consuming, but typically less than the upper bound.

\subsection{Min put.}\label{minput}
In this example, the state variable is $X_t \equiv x_t$, the vector of discounted log prices, and  the reward function for stopping at time $t$ will be
\begin{equation}
g(t,X_t)  = \bigl(\; Ke^{-rt} - \exp( \min_{1 \leq i \leq d} X^i_t)\;  \bigr)^+.
\label{minputg}
\end{equation}
This example was studied in \cite{R1}, and the figures in the column MC price of Table \ref{Table1} were taken from that paper; for $d= 30, \, 60$ no values are given in \cite{R1}. The column headed {\it European} is the simulation value for the European option, where no early exercise is allowed. The columns headed {\it low} and {\it high} are sample means obtained from a simulation method.  Standard errors are reported in brackets after the mean values. Notice that the lower bound is in all cases less good than the bound from the European price. This may be in part due to the finite sizes of the bins, and the error arising from that, but it indicates that this simple approach is not able to extract the holder's  early exercise value of the option.  Nevertheless, the bounds are reasonably close for all values of $d$, even quite good for larger values of $d$, and the run times are going up roughly linearly with $d$.

\begin{table}[ht] 
\begin{center}
\begin{tabular}{|c|c|c|c|c|c|c|}
\hline 
$d$ & European &low & MC price & high & gap(\%) & time \\ 
\hline 
2 & 24.78 (0.07) & 24.71 (0.08) & 25.16 & 25.65 (0.30) & 3.40 & 8.89 \\
\hline 
3 & 31.27 (0.06) & 31.16 (0.08) & 31.76 & 32.29 (0.34) & 3.16 & 10.20 \\
\hline 
4 &  35.75 (0.06) & 35.74 (0.07) & 36.28 & 36.81 (0.35) & 2.88 & 11.42 \\ 
\hline  
5 & 39.22 (0.06) & 39.12 (0.07) & 39.47 & 39.84 (0.33) & 1.56 & 12.49 \\
\hline 
10  & 48.01 (0.04) & 47.81 (0.05) & 48.33 & 48.50 (0.28) & 1.00 & 18.64  \\
\hline 
15   & 52.13 (0.04) & 52.06 (0.04) & 52.14 & 52.62 (0.23) & 0.93 & 26.38   \\
\hline 
30   & 57.82 (0.03) & 57.66 (0.03) & -  & 58.21 (0.17) & 0.66 & 61.20  \\
\hline
60 & 62.28 (0.02) & 62.18 (0.03) & - & 62.57 (0.16) & 0.47 & 183.49
\\
\hline
\end{tabular}
\caption{Min put prices. The $d$ assets are independent, $S^i(0)=100$.
Other parameters are $K=100$, $r=0.06$, $T=0.5$, $\sigma_{ii} = 0.6$. Parameters for the simulations are  $\Nbins =   200, \Nblock =  200, N_T =   40, \Nprimal =    50000, \Ndual =   400,  \Nsub =   60$.}
\label{Table1}
\end{center}
\end{table}

\subsection{Max call.}\label{maxcall}
As with the min put, the state variable is $X_t \equiv x_t$, the vector of discounted log prices, but this time  the reward function for stopping at time $t$ will be
\begin{equation}
g(t,X_t)  = \bigl(\;  \exp( \max_{1 \leq i \leq d} X^i_t)\; - Ke^{-rt}
 \bigr)^+.
\label{maxcallg}
\end{equation}
In order to make the problem interesting, the assets will be assumed to pay dividends at a constant rate.  This example was studied by Broadie \& Glasserman\footnote{In the
preprint version, the example presented in Table \ref{Table2} is said to give results for common expiry $T=3$ with 3, 6, 9 equally-spaced exercise opportunities, but the published version gives the same numerical values purportedly for end-of-year exercise opportunities with expiries 3, 6, 9 years. Our numerics show that in fact the problem solved is correctly stated in the preprint, and mis-stated in the published version. }\cite{BG}, and has been used as a test example in a number of other studies, including \cite{FLMSW}, \cite{HK}, \cite{AB}, \cite{JLTW}. 
 Notice that this time the lower bound we obtain is significantly bigger than the European price, so the holder is able to use the simple Markov coercion heuristic to extract some of the early exercise value. The bounds are less close than for the min put example, but this is a more difficult option to handle; errors of the order of 5-6\% need to be improved, but already give usable information.

\begin{table}[ht]
\begin{center}
\begin{tabular}{|c|c|c|c|c|c|c|c|}
\hline 
$m$ & $S_0$ & European & low & BG price & high & gap(\%) & time \\ 
\hline 
	&90 & 14.62 (0.06) & 15.39 (0.08) & 16.006 & 16.18 (0.10) & 4.87 & 4.40 \\
3 & 100  & 22.98 (0.08) & 24.37 (0.09) & 25.284 & 25.43 (0.12) & 4.15 & 4.38  \\
  &110 & 32.60 (0.09) & 33.93 (0.11) & 35.695 & 35.86 (0.14) & 5.39 & 4.49 \\
\hline 
     &  90 & 14.58 (0.06) & 15.92 (0.07) & 16.474 & 16.65 (0.08) & 4.39 & 8.00\\
6  & 100 & 23.14 (0.08) & 24.95 (0.08) & 25.290 & 26.31 (0.10) & 5.17 & 8.35 \\
 & 110 & 32.64 (0.09) & 35.03 (0.09) & 36.479 & 37.24 (0.13) & 5.94 & 8.30   \\
\hline 
  & 90  & 14.57 (0.06) & 16.05 (0.07) & 16.659 & 16.93 (0.09) & 5.20 & 11.77  \\
9 & 100  & 23.05 (0.08) & 25.14 (0.08) & 26.158 & 26.88 (0.11) & 6.47 & 12.08 \\
 & 110 & 32.61 (0.09) & 35.23 (0.09) & 36.782 & 37.57 (0.12) & 6.23 & 11.83
 \\
\hline 
\end{tabular} 
\caption{Max call prices on 5 independent assets with common volatility $\sigma = 0.2$ and expiry $T=3$. There are $m=3, 6, 9$ exercise opportunities at times $iT/m, \; i=0,\ldots,m$.  Other parameters are $K=100$, $r = 0.05$, $\delta = 0.1$.   Simulation parameters are
 $\Nbins =   500, \Nblock =  100,  \Nprimal =   50000, \Ndual =  4000,  \Nsub =  150$
.}
\label{Table2}
\end{center}
\end{table}

\subsection{Basket put.}\label{basketput}
This is an example studied in Kovalov, Linetsky \& Marcozzi \cite{kovalov}, and subsequently in Jin {\it et al} \cite{JLTW}.  The state variable is again the vector of $d$ discounted log prices, and this time the stopping reward function is
\begin{equation}
g(t,X_t)  = \bigl(\;Ke^{-rt}  -  d^{-1} \sum_{i=1}^d \exp( X^i_t)\; 
 \bigr)^+.
\label{basketg}
\end{equation}
All the stocks start  at 100, the strike  is $K=100$,  the riskless rate is $0.03$, the expiry is $T=0.25$, and the individual asset volatilities are all 0.2, but this time the assets are not supposed independent; there is constant correlation $\rho = 0.5$ between all the assets\footnote{The specification of the various parameters in \cite{kovalov} is internally inconsistent, and not in agreement with the parameters quoted in \cite{JLTW}. For the reported prices to be correct, we find that the parameter values used must be those we have stated.}. Kovalov {\it et al} use a numerical PDE approach, Jin {\it et al} use a simulation methodology, and both approaches appear to give better precision than the method we have used here. Nevertheless, as we shall soon see, the difference in precision is not practically relevant.  We  report the results in Table \ref{Table3}.
\begin{table}[ht]
\begin{center}
\begin{tabular}{|c|c|c|c|c|c|c|}
\hline 
$d$ & European & low & KLM & high & gap(\%) & time \\ 
\hline  
2 & 3.08 (0.01) & 3.13 (0.02) & 3.14 & 3.25 (0.01) & 3.59 & 41.89 \\
\hline  
3   & 2.89 (0.01) & 2.93 (0.02) & 2.94 & 3.04 (0.01) & 3.64 & 48.03     \\
\hline  
4   & 2.78 (0.01) & 2.81 (0.02) & 2.84 & 2.94 (0.01) & 4.35 & 52.39   \\
\hline 
5  & 2.72 (0.01) & 2.75 (0.02) & 2.77 & 2.87 (0.01) & 3.99 & 57.54   \\ 
\hline 
6  & 2.68 (0.01) & 2.73 (0.01) & 2.73 & 2.83 (0.01) & 3.81 & 63.42   \\
\hline
12  & 2.56 (0.01) & 2.60 (0.01) & - & 2.70 (0.01) & 3.56 & 99.03  \\
\hline 
\end{tabular}  
\caption{Basket put. All assets start at 100, $K=100$, $T=0.25$, $r=0.03$. All assets have volatility $20\%$, and the correlation between assets is 0.5. 
Other parameters are 
$\Nbins =   500, \Nblock =  200, N_T =   40, \Nprimal =    50000, \Ndual =  1000,  \Nsub =  160$. }
\label{Table3}
\end{center} 
\end{table}

These values were computed assuming that the volatility parameter $\sigma$ is equal to 20\%. But are we sure of that? In any application, the volatility (assumed constant) would have to be estimated; are we really sure that the volatility is not 19\%?   Or 21\%? Are we really sure that the volatility will remain constant at 20\% until expiry of the option?  Suppose we repeat the calculations of Table \ref{Table3} for those values of the volatility parameter and see what ranges for the price result.  The outcomes are recorded in Table \ref{Table4}, and what we see is that {\em there is  no overlap between the computed intervals for the price for the three values of $\sigma$}.  In other words, {\em the uncertainty in the price arising from our simulation bounds is comparable to the uncertainty in price which would arise from the uncertainty in the input parameter values.}
\begin{table}[ht]
\begin{center}
%

\begin{tabular}{|c|c|c||c|c||c|c|}
\hline 
$d$ & low (19\%) & high (19\%) & low (20\%) & high (20\%) & low (21\%) & high (21\%) \\ 
\hline 
2 & 2.98 & 3.07 &3.13 &3.25 & 3.28 & 3.42 \\ 
\hline 
3 &2.76 & 2.87 &2.93  & 3.04 &3.09  & 3.20 \\ 
\hline 
4 & 2.68  & 2.79 &  2.81 & 2.94 & 2.97 & 3.10 \\ 
\hline 
5 & 2.61 & 2.72 & 2.75 & 2.87 & 2.93&  3.04 \\ 
\hline 
6 & 2.58 & 2.68  & 2.73 & 2.83  & 2.86&  2.97\\ 
\hline 
\end{tabular}

\caption{Basket put. Parameters as for Table \ref{Table3}, except the volatility parameter which takes values 19\%, 20\%, 21\%.  }
\label{Table4}
\end{center} 
\end{table}

\subsection{Fixed strike Bermudan-Asian call.}\label{FixedBAcall}
In this example, there is a single asset $S$, and the reward for stopping at time $\tau$ is
\begin{equation}
 g(\tau,X_\tau) = e^{-r\tau}( A_\tau - K)^+,
 \label{asian_g}
 \end{equation}
where we define the average price
\begin{equation}
A_t	= \frac{\int_{-\delta}^t S_u \; du}{t + \delta}.
\label{Adef}
\end{equation}
Here, $\delta >0$ is some initial window required to prevent wild oscillations. There is also some initial lock-out time $t^* \geq 0$ during which exercise of the option is forbidden. The state variable of the problem is $X_t = [ S_t, A_t, t]$.  The numerical results are reported in Table \ref{Table5}.  The  gaps between the bounds are quite variable.  In fact we used three different numeraires (the bank account, the stock, and the martingale $E_t[A_T]$).  The results are usable, but not particularly good, and this is really because this example is intrinsically two-dimensional, so any attempt to coerce it to one dimension is missing something essential. We cannot expect a stopping rule which only looks at $A_t$ to do very well, because the current value of $S_t$ has to be considered as well; if $S_t$ is high enough, the value $g(t,X_t)$ is actually increasing, so we would certainly  not stop at such a time. But a rule that only considers $A_t$ would not understand that.

\begin{table}[ht]
\begin{center}

\begin{tabular}{|c|c|c|c|c|c|c|c|}
  \hline 
  $A_0$ & $S_0$ & European & low & FD price & high & gap(\%) & time \\ 
  \hline 
  & 80 & 0.956 (0.008) & 0.945 (0.012) & 0.949 & 0.974 (0.033) & 1.83 & 29.04\\
  \hline 
& 90 & 3.230 (0.016) & 3.216 (0.023) & 3.267 & 3.374 (0.073) & 4.27 & 29.51\\
  \hline 
90 & 100 & 7.571 (0.025) & 7.568 (0.035) & 7.889 & 8.048 (0.116) & 5.93 & 29.39\\
  \hline 
 & 110 & 13.78 (0.03) & 13.77 (0.04) & 14.538 & 14.81 (0.16) & 6.97 & 29.72\\
  \hline 
 & 120 & 21.21 (0.03) & 21.00 (0.05) & 22.423&  22.94 (0.18) & 7.58 & 30.01\\
  \hline \hline
   & 80 & 1.098 (0.009) & 1.088 (0.013) & 1.108 & 1.160 (0.036) & 5.36 & 29.39\\
  \hline 
   & 90& 3.557 (0.017) & 3.578 (0.024) & 3.710 & 3.759 (0.080) & 4.82 & 29.40 \\ 
  \hline 
  100 & 100& 8.133 (0.025) & 8.148 (0.036) & 8.658 & 8.905 (0.127) & 8.50 & 29.40\\ 
  \hline 
   & 110 & 14.73 (0.04) & 13.92 (0.05) & 15.717 & 16.39 (0.17) & 10.12 & 29.62\\
  \hline 
   & 120 & 22.09 (0.05) & 21.88 (0.07) & 23.811 & 24.46 (0.21) & 9.67 & 27.32\\
  \hline \hline
   & 80 & 1.260 (0.011) & 1.233 (0.015) & 1.288 & 1.337 (0.048) & 5.75 & 30.32 \\
  \hline 
   & 90 & 3.911 (0.023) & 3.978 (0.033) & 4.136 & 4.506 (0.107) & 11.72 & 27.19 \\
  \hline 
  110 & 100 & 8.885 (0.029) & 8.378 (0.040) & 9.821 & 10.822 (0.133) & 17.90 & 29.64\\
  \hline 
   & 110 & 15.53 (0.04) & 15.61 (0.05) & 17.399 & 18.40 (0.16) & 15.14 & 29.69\\
  \hline 
   & 120 & 23.02 (0.05) & 23.36 (0.07) & 25.453 & 26.42 (0.20) & 10.92 & 26.94 \\
  \hline 
  \end{tabular}

\caption{Fixed strike Bermudan Asian call. Parameters are $\sigma = 0.2$, $K=100$, $t^* = 0.25$, $\delta = 0.25$, $T=2$.  Other parameters are
$\Nbins =   500, \Nblock =  100, N_T =   40, \Nprimal =   50000, \Ndual =  4000,  \Nsub =  125$.}
\label{Table5}
\end{center} 
\end{table}

\subsection{Floating strike Bermudan Asian call}\label{FloatingBAcall}
The story is very similar to Section \ref{FixedBAcall}, except that the reward  is 
\begin{equation}
 g(\tau,X_\tau) = e^{-r\tau}( A_\tau - S_\tau)^+
 \label{asian_g_2}
 \end{equation}
 for stopping at time $\tau$. This example is in fact much easier than the fixed strike, because the process $A_t/S_t$ is a Markov process already. Scaling says we need only vary $S_0$ while keeping $A_0$ fixed, which is what we do.  The results obtained  when we fix $A_0=100$ are given in Table \ref{Table6}. Once again, the gaps between the two bounds are usably close, and the lower bounds are well clear of the European values, so here the approximate stopping rule gives a very substantial improvement.
\begin{table}[ht]
\begin{center}
\begin{tabular}{|c|c|c|c|c|c|}
\hline 
$S_0$ & European& low & high & gap(\%) & time \\ 
\hline 
80& 3.98 (0.04) & 10.62 (0.03) & 10.81 (0.07) & 1.77 & 8.88\\ 
\hline
90 & 3.936 (0.041) & 8.347 (0.032) &  8.618 (0.051) & 3.14 & 9.21 \\
\hline 
100  & 3.866 (0.042) & 7.136 (0.033) &  7.485 (0.040) & 4.66 & 9.11 \\
\hline 
110    & 3.873 (0.043) & 6.558 (0.035) & 6.909 (0.043) & 5.09 & 8.93   \\
\hline 
120   & 3.942 (0.046) & 6.137 (0.037) &  6.529 (0.036) & 6.00 & 8.82  \\
\hline 
\end{tabular}  
\caption{Floating strike Bermudan Asian call. Parameters are $\sigma = 0.2$, $A_0=100$, $t^* = 0.25$, $\delta = 0.25$, $T=2$.  The calculations were done in the numeraire of the discounted asset price.
Other parameters are  $\Nbins =   200, \Nblock =  100, N_T =   40, \Nprimal =   50000, \Ndual =  4000,  \Nsub =   50$
. }
\label{Table6}
\end{center} 
\end{table}

\subsection{Fixed window lookback option.}\label{range}
This example illustrates the capacity of the methodology to handle high-dimensional problems. Here we suppose that the stock price is recorded at times which are multiples of some $h>0$, and stopping at time $\tau = kh$ delivers reward
\begin{equation}
g(\tau,X_\tau) = \sup_{k-a \leq j \leq k} S_{jh}  ,
\label{lookback_g}
\end{equation}
where $a$ is some positive integer.  This time, the state variable $X$ has to record the last $a$ values of the price, since the sup and inf are taken over a fixed window. The results are presented in Table \ref{Table7}.  Notice that for this example the bounds are very close, getting slightly less good as the lookback parameter $a$ rises. Equally noteworthy is the fact that the run times are changing little as we increase the lookback parameter; so in the final row of the table, the state variable is 25-dimensional. It should not be a surprise that there is so little variation in run times, because increasing $a$ makes no difference to the simulation load;  each period, we simulate one new value, all that is different is that we are storing more or fewer values from the past.
\begin{table}[ht]
\begin{center}

\begin{tabular}{|c|c|c|c|c|c|}
\hline 
$a$ (days) & European & low & high & gap(\%) & time \\ 
\hline 
5 & 103.50 (0.02) & 117.67 (0.01) & 118.51 (0.03) & 0.71 & 93.00\\
\hline 
10 & 105.96 (0.03) & 124.71 (0.02) & 126.46 (0.05) & 1.38 & 88.65 \\
\hline 
15& 107.80 (0.04) & 129.24 (0.03) &  131.86 (0.06) & 1.98 & 91.34\\
\hline 
20  & 109.38 (0.05) & 132.62 (0.03) &  135.74 (0.07) & 2.30 & 96.68\\
\hline 
25 & 110.79 (0.06) & 135.35 (0.04) & 138.77 (0.07) & 2.47 & 100.58\\
\hline 
\end{tabular} 
\caption{Fixed window lookback option. Parameters were $T=1$, $\sigma = 0.5$, $r=0.05$,
$S_0=100$, and the time interval was divided into 500 equal  (half-day) time steps. The calculations were done in the numeraire of the discounted asset price.
Other parameters were
$\Nbins =   250, \Nblock =   60, N_T =  500, \Nprimal =   50000, \Ndual =  4000,  \Nsub =   25$.}
\label{Table7}
\end{center} 
\end{table}

\subsection{Fixed window range option.}\label{FixedRange}
This example is similar to the fixed window lookback option of Section \ref{range}, except that the reward for stopping at time $\tau = kh$ is 
\begin{equation}
g(\tau,X_\tau) = \sup_{k-a \leq j \leq k} S_{jh} -  \inf_{k-a \leq j \leq k} S_{jh},
\label{range_g}
\end{equation}
where $a$ is some positive integer.  The results are reported in Table \ref{Table8}.
The gaps between the upper and lower bounds are higher than for the lookback example expressed as a percentage, but the arithmetic gaps are roughly comparable, with similar run times.
\begin{table}[ht]
\begin{center}

\begin{tabular}{|c|c|c|c|c|c|}
\hline 
$a$ (days) & European & low & high & gap(\%) & time \\ 
\hline 
5& 9.94 (0.03) & 23.97 (0.02) & 25.35 (0.04) & 5.43 & 31.56\\
\hline 
10 & 16.77 (0.05) & 33.31 (0.03) & 36.63 (0.07) & 9.05 & 35.85\\
\hline 
15 & 21.94 (0.06) & 39.56 (0.03) &  44.63 (0.09) & 11.36 & 38.05\\
\hline 
20& 26.25 (0.07) & 44.36 (0.04) & 51.00 (0.11) & 13.01 & 42.28\\  
\hline
25& 30.04 (0.08) & 48.39 (0.04) & 56.34 (0.14) & 14.12 & 45.00\\ 
\hline 
\end{tabular} 
\caption{Fixed window range option. Parameters were $T=1$, $\sigma = 0.5$, $r=0.05$,
$S_0=100$, and the time interval was divided into 250 equal time steps. The calculations were done in the numeraire of the discounted asset price.
Other parameters were
$\Nbins =   200, \Nblock =   50, N_T =  250, \Nprimal =   50000, \Ndual =  4000,  \Nsub =   25$.}
\label{Table8}
\end{center} 
\end{table}

\subsection{Min puts with stochastic volatility and interest.}\label{SISV}

In this example, we consider a situation where there are $d>1$ assets, with stochastic volatility and interest rates. There are examples of such models in various places in the literature, for example, Medvedev \& Scaillet \cite{MS}, Boyarchenko \& Levendorskii \cite{BL}, Jin {\it et al.} \cite{JLTW}. Heston dynamics for the asset and the volatility are popular in theoretical work, but it is far from clear that the dependence of the volatility of volatility on level takes the square-root form postulated in the Heston model; and still less is it persuasive that the Cox-Ingersoll-Ross interest-rate model correctly describes the volatility of interest rates when those rates are low.  As this is a simulation study, we are freed from any need to choose models that are theoretically tractable\footnote{Tractability is in any case illusory; we regard a model as tractable if there is a closed-form solution for {\em a small number} of derivative prices, overlooking the fact that for the majority of derivative prices there is no closed form solution.}, so we may make some modelling assumptions that match observed behaviour better.

So our story supposes that there is some market Brownian motion $W^M$, and that 
log prices $x^i_t \equiv \log S^i_t$ evolve as
\begin{equation}
dx_t^i = \sigma^i_t \{ \,  \rho_S\, dW^M_t + \rho'_S\, dW^{S,i}_t\, \} +(r_t - \half (\sigma_t^i)^2 )\, dt
\label{dx_i}
\end{equation}
where $r$ is the riskless rate process, and $\rho_S \in (0,1)$ is the correlation\footnote{When $\rho$ is a correlation coefficient, we use $\rho'$ to denote $\sqrt{1-\rho^2}$.} of the log prices with the market Brownian motion.  The process $W^S$ is a $d$-dimensional Brownian motion independent of $W^M$.

The volatility process $\sigma$ appearing in \eqref{dx_i} is represented as
\begin{equation}
  \sigma^i_t  = \bar\sigma^i \exp( \, \xi^i_t\, )
  \label{sigmai}
\end{equation}
in terms of constants $\bar\sigma^i$ and a process $\xi$  which is an OU process evolving as 
\begin{equation}
d\xi_t =  -\beta_\xi\,  \xi_t \,  dt + \sigma_\xi \, (\rho_\xi \, dW^M_t
+\rho'_\xi\, dW^\xi_t \, ).
\label{dxi}
\end{equation}

Finally, our model for the interest-rate process $r$ is just a Black-Karasinski model: we have $r_t = \bar{r}\exp(z_t)$, where 
\begin{equation}
dz_t  = - \beta_r \, dt +\sigma_r\, (\rho_r\, dW^M_t + \rho'_r \, dW^r_t)
\label{dr}
\end{equation}
for some constants $\beta_r>0$, $\sigma_r>0$ and $\rho_r$, which would typically be assumed positive since we expect that as the market rises the rate of interest should also rise.

Altogether then, this is a simple but sprawling model; even assuming (as we do here) that correlations are common across stocks, the parameter vector is
\begin{equation}
\theta = ( \,  \rho_S, \rho_\xi, \rho_r, (\bar\sigma^i), \bar{r}, \beta_\xi,\sigma_\xi,
\beta_r, \sigma_r).
\label{thetadef}
\end{equation}
What would be reasonable values for these parameters?  For the interest rate evolution, we shall be guided by Black \& Karasinski \cite{BK} and take $\bar r = 0.06$, $\sigma_r = 0.12$,  $\beta_r = 0.02$, and $\rho_r = 0.3$. Correlations between stocks are variable, but typically in the range 0.25-0.60; we shall take $\rho_S = 0.3$. For simplicity we assume all stocks have common volatility $\bar \sigma^i  = 0.6$.  Fluctuations in volatilities are of the order of tens of percent, so by comparing with the standard deviation of an OU process, we impose
\begin{equation}
\frac{\sigma_\xi}{\sqrt{2 \beta_\xi}}  = 0.1.
\label{setpars}
\end{equation}
We shall set $\beta_\xi = 4.5$, so that the mean reversion time for volatility is of the order of three months, and this gives from \eqref{setpars} that $\sigma_\xi = 0.3$. Finally, we take $\rho_\xi = 0.3$.

For the rest of our discussion, we shall focus on the case where $d=5$, that is: there are five assets. We will also restrict attention to min puts, taking strike $K=100$ and all assets starting at 100 throughout. We shall also assume that the expiry $T=0.5$.  This will allow us to investigate the effects of varying initial values of $r$ and $\sigma$, as well as various parameters.  If we make $\rho_S=0$ and make the mean reversion parameters $\beta_r$, $\beta_\xi$ very large, we have in effect got back to the situation with independent assets,  constant volatilities and interest rate that was studied in Examples \ref{minput}, \ref{maxcall}.  So we should see the same answer; and we do - the range from this calculation came out to be [39.06, 40.65], which is the same as we found in Table \ref{Table1}, to within sampling error.

Next, we can see what happens when we keep the volatility and interest rate constant, but allow $\rho_S$ to vary; the results are in Table \ref{Table9}.  What we see is that while the correlation is not too far from zero, there is no clear effect on the price, but as the correlation between the assets rises, the price of the min put falls. This is to be expected; the higher the correlation the more alike the assets are, so there will be less dispersion in the prices at any time, so the minimum will be higher.
\begin{table}[ht]
\begin{center}

\begin{tabular}{|c|c|c|c|c|c|}
 \hline 
 $\rho_S$ & European& low & high & gap (\%) & time \\ 
 \hline 
 -0.15 & 38.81 (0.11) & 38.81 (0.07) &  39.95 (0.16) & 2.85 & 19.87\\
 \hline 
 0.00 & 39.32 (0.11) & 39.09 (0.07) &  40.16 (0.16) & 2.09 & 20.10\\
 \hline 
 0.15& 38.85 (0.11) & 38.72 (0.07) & 39.90 (0.16) & 2.63 & 20.11\\
 \hline 
 0.30 & 37.58 (0.12) & 37.63 (0.07) & 38.87 (0.17) & 3.17 & 20.16\\
 \hline 
 0.45 & 36.00 (0.13) & 35.82 (0.08) & 37.06 (0.19) & 2.88 & 20.15\\
 \hline 
 0.60 & 33.33 (0.13) & 33.00 (0.08) &  34.15 (0.20) & 2.40 & 20.12 \\
  \hline
 \end{tabular}  
\caption{Prices of min puts as $\rho_S$ varies. Volatility is constant at $\bar\sigma = 0.6$, interest is constant at $r = 0.06$.
Other parameters are
 $\Nbins =   200, \Nblock =   50, N_T =   40, \Nprimal =   50000, \Ndual =  4000,  \Nsub =   50$.  }
\label{Table9}
\end{center} 
\end{table}

Now we relax the assumption of constant interest rate, and let the interest rate evolve as in the Black-Karasinski specification \eqref{dr}, fixing $\rho_S$ at its default value 0.3, and observing the effects of different initial values of the riskless rate. We hold the volatility constant at the default value 0.6. The results are reported in Table \ref{Table10}. As the initial interest rate rises\footnote{The row of the table for $r_0=0$ was obtained by taking $\log r_0 = -15.6$.}, the price of the min put falls, as would be expected from the stronger discounting of the stopping reward; even allowing for the fact that we can only obtain an interval for the price, the effect of change of initial interest rate is discernible.
\begin{table}[ht]
\begin{center}

\begin{tabular}{|c|c|c|c|c|c|}
 \hline 
 $r_0$ & European & low & high & gap (\%) & time \\ 
 \hline 
 0.00 & 40.61 (0.12) & 40.45 (0.07) & 41.54 (0.18) & 2.26 & 19.64  \\
 \hline 
 0.025 & 39.34 (0.12) & 39.27 (0.07) & 40.16 (0.18) & 2.03 & 19.58\\
 \hline 
 0.06 & 37.76 (0.12) & 37.50 (0.07) &  38.70 (0.17) & 2.42 & 20.04\\
 \hline 
 0.10 & 35.88 (0.12) & 35.86 (0.07) & 36.99 (0.17) & 2.99 & 19.68  \\
 \hline 
 \end{tabular}  
\caption{Prices of min puts as $r_0$ varies. Volatility is constant at $\bar\sigma = 0.6$, correlation between assets is $\rho_S = 0.3$.
Parameters are $ \rho_r = 0.3$,  
$\bar r = 0.06$, $\beta_r = 0.02$, $\sigma_r = 0.12$.
Other parameters are
 $\Nbins =   200, \Nblock =   50, N_T =   40, \Nprimal =   50000, \Ndual =  4000,  \Nsub =   50$.   }
\label{Table10}
\end{center} 
\end{table}

Having seen the effect of changing initial interest rate while volatility is held constant, we next hold the interest rate constant at its default value $\bar r = 0.06$, and allow the volatility to be stochastic. The impact of different initial levels of volatility is shown in Table \ref{Table11}, where it is assumed that the initial volatility is common across all the assets. Again, we see quite pronounced effect of the initial volatility on the price of the min put option; the price rises as the initial volatility increases, as one would expect.
\begin{table}[ht]
\begin{center}

\begin{tabular}{|c|c|c|c|c|c|}
 \hline 
 $\sigma_0$ & European & low & high & gap (\%) & time \\ 
 \hline
 0.10  & 22.17 (0.09) & 22.14 (0.05) &  22.89 (0.09) & 3.15 & 19.67\\
 \hline 
 0.20 & 26.78 (0.10) & 26.66 (0.06) &  27.42 (0.12) & 2.34 & 19.98 \\
 \hline 
 0.30  & 30.08 (0.10) & 29.99 (0.06) &  31.03 (0.13) & 3.04 & 19.97 \\
 \hline 
 0.40 & 32.83 (0.11) & 32.86 (0.07) & 33.74 (0.15) & 2.59 & 20.08 \\
 \hline 
 0.50 & 35.37 (0.11) & 35.33 (0.07) &  36.40 (0.16) & 2.84 & 20.09 \\
 \hline 
 0.60 & 37.49 (0.12) & 37.50 (0.07) &  38.65 (0.17) & 2.97 & 20.10\\
 \hline 
 \end{tabular}  
\caption{Prices of min puts as $\sigma_0$ varies. Interest is constant at $\bar r = 0.06$,  correlation between assets is $\rho_S = 0.3$.
Parameters are $ \rho_\xi= 0.3$, $\bar\sigma^i = 0.6$ for all $i$, 
$\beta_\xi = 4.5$, $\sigma_\xi = 0.3$. Other parameters are
 $\Nbins =   200, \Nblock =   50, N_T =   40, \Nprimal =   50000, \Ndual =  4000,  \Nsub =   50$.
 }
\label{Table11}
\end{center} 
\end{table}

The final study considers the full model where both volatility and interest rate are stochastic.  There are too many parameters to explore in a paper, so we content ourselves with holding the parameters fixed at their default values, and varying the initial riskless rate and initial volatility. We see the results in Table \ref{Table12}. Again the comparative statics behave as one would expect, with the magnitudes of the effects of changing initial values being big enough to show up clearly,
even allowing for the fact that we have only got bounds on the prices.

\begin{table}[ht]
\begin{center}

\begin{tabular}{|c|c|c|c|c|c|c|}
  \hline 
  $r_0$ & $\sigma_0$ & European& low & high & gap (\%) & time \\ 
  \hline 
   & 0.2 & 29.71 (0.10) & 29.53 (0.06) &  30.38 (0.12) & 2.18 & 19.47\\
  0.00 & 0.4 & 35.77 (0.11) & 35.65 (0.07) & 36.45 (0.15) & 1.86 & 19.64 \\
  
   & 0.6 & 40.38 (0.12) & 40.36 (0.07) &  41.34 (0.17) & 2.32 & 19.73\\
  \hline 
   & 0.2 & 26.85 (0.10) & 26.69 (0.06) & 27.61 (0.12) & 2.74 & 19.84\\

  0.06 & 0.4 & 33.00 (0.11) & 32.78 (0.07) & 33.60 (0.15) & 1.79 & 20.20\\
  & 0.6 & 37.67 (0.12) & 37.70 (0.07) &  38.66 (0.17) & 2.49 & 20.00\\
  \hline 
  \end{tabular}   
\caption{Prices of min puts as $\sigma_0$ and $r_0$ vary.
Parameters are $\rho_S= \rho_\xi = \rho_r = 0.3$, $\bar\sigma^i = 0.6$ for all $i$, 
$\bar r = 0.06$, $\beta_\xi = 4.5$, $\sigma_\xi = 0.3$, $\beta_r = 0.02$, $\sigma_r = 0.12$.
Other parameters are 
 $\Nbins =   200, \Nblock =   50, N_T =   40, \Nprimal =   50000, \Ndual =  4000,  \Nsub =   50$. }
\label{Table12}
\end{center} 
\end{table}

\section{Conclusions and discussion.}\label{C+D}
The aim of this paper has been to see to what extent we are able to solve the problem of pricing Bermudan options in very high dimensions.   Once we accept that for such problems it is impossible that we can know the value function, we realize that in fact various approaches which have been developed in the last twenty or so years may be combined to provide a practical solution in many instances.  Working entirely with Markovian problems, the key elements to the approach studied here are:
\begin{itemize}
\item pretend that the stopping reward process $Z$ is itself Markovian, and by discretizing $Z$ onto a suitably-chosen finite set of values we estimate the transition probabilities of this finite state Markov chain by simulation ( this is the approach of Barraquand \& Martineau \cite{BM});
\item solve the optimal stopping problem for this finite state Markov chain by dynamic programming;
\item use the solution to generate a stopping rule whose performance is evaluated by simulation;
\item use the dual characterization of the value of the problem (see \cite{R1}, \cite{HK}, \cite{AB}) to find a hedging martingale. 
\end{itemize}
This approach is a very general methodology that delivers upper and lower bounds on the price that are generally reasonably close, but more importantly it provides effective recipes for action. It is often said that the price a bank charges for a derivative is more to do with the cost of hedging that derivative than with any number that comes out of some model; and the approach we advocate here puts that into effect. Indeed, at each time the analysis we propose tells the seller of the option what hedge he should use - all he has to do is to hedge the approximate value at the next time step. That approximate value is a simple function of the stopping reward at the next time step.  Similarly, the approach we use provides a compact solution for the buyer of the option; at each time, he calculates the value of immediate stopping, and stops if and only if this value is in some finite union of intervals. 

This approach is remarkably successful in many of the examples we have studied, providing bounds which are often within 5\% of each other. Since this sort of error is inherent in the estimates of input parameters, or in the assumption that those parameters are constant over time, there is really little benefit in getting the bounds very much tighter. We would ideally like to have methods which (if they cannot get the derivative price exactly) can give bounds which are apart by, say, 1 basis point.  This is an industry-standard criterion ... but where does it come from? Does a bank {\em really} care if their calculation of the price of a derivative is out by 1 cent in \$100? Of course not!  The 1bp  criterion really comes from the desire to delta-hedge the option; so we want to vary the prices of the underlyings by $\pm$1\% and then find the change in the price in order to put on the delta hedge, and at this point 1bp accuracy is a relevant requirement.  But the approach here gives us the hedging strategy by a completely different route - there is {\em no} delta hedging, only the hedge that comes from the dual approach!  Moreover, getting 1bp accuracy from a simulation method is already rather over-optimistic.  

So what happens when the upper and lower bounds are further apart, as in the fixed-strike Bermudan Asian option?  This is not problematic conceptually; the lower bound is what the buyer objectively thinks the option {\em on its own} is worth, the upper bound is objectively what the seller expects hedging this derivative {\em on its own} to cost him, but either side may move beyond the bound if by taking on the contract they lay off risk elsewhere in their portfolio - this is the most basic reason for a market in derivatives. There is nothing difficult or contradictory in a market where the bid price is below the ask price - this is normal!   If we find a situation where the bounds are far apart, and we feel it is important to bring the bounds closer together, what can we do?  There are four places where error enters into this approach:
\begin{enumerate}
\item the assumption that the reward process $Z$ is a real-valued Markov process;
\item the error from discretizing $Z$ values into bins, and deriving the transitions from simulation;
\item simulation error in evaluating the stopping strategy;
\item simulation error in evaluating the hedging strategy.
\end{enumerate}
We can reduce the last two by doing many more simulations, and the second can also be addressed by taking more bins and doing more simulations, but {\em the first error is intrinsic} - we can do nothing to reduce it other than change the problem in some way. The fixed-strike Bermudan Asian option illustrates this well, because as we discussed in Example \ref{FixedBAcall}, the state variable for this particular problem really has to be the two-dimensional vector $(S_t, A_t)$, and the crude notion that we can get a good approximation just by looking at $A_t$ on its own is shown by our calculations to be wide of the mark. Now of course we could work to exploit the specific features of this problem to devise a problem-specific solution (and the finite-difference calculations in Longstaff \& Schwartz \cite{LongstaffSchwartz} and Rogers \& Shi \cite{RogersShi} are instances), but this is contrary to the {\em generic} nature of the approach presented here. If we wanted to continue to use this approach, we might try to bin the values of the bivariate process $(S_t, A_t)$, which is of course conceptually no different from the binning of the scalar-valued reward process, but we may expect that the coarser binning we can expect from a two-dimensional underlying process will raise the second type of error.

To summarize, then, the approach explored in this paper:
\begin{itemize}
\item is completely generic - the same code gets used with only changes in the Markov process and the reward function;
\item always gives bounds, which are often within the range of error introduced by estimation or questionable modelling assumptions - and always within the profit margin required for an OTC product;
\item is largely insensitive to dimension;
\item gives simple and explicit exercise strategies, and hedging strategies;
\item requires only the ability to simulate a step of the underlying Markov process.
\end{itemize}
The approach of this study puts together earlier discoveries from the last twenty years. While it may be premature to declare that Bermudan options in high dimensions are now a solved problem, what we have seen is that there is a general approach which proves to be at very least a good start on the problem, even if for a particular question we may want to dig deeper. Once we set the question in the context of estimation error and model uncertainty, it may indeed seem pointless to dig deeper in any case.

\pagebreak
\bibliography{BOBS}
\bibliographystyle{plain}

\end{document}